\documentstyle[12pt]{article}
\newcommand{\bk}{{\bf k}}
\newcommand{\br}{{\bf r}}

\newcommand{\bR}{{\bf R}}
\newcommand{\bP}{{\bf P}}

\newcommand{\bX}{{\bf X}}
\newcommand{\bx}{{\bf x}}
\newcommand{\by}{{\bf y}}
\newcommand{\bp}{{\bf p}}
\newcommand{\bq}{{\bf q}}

\newcommand{\hk}{{\hat{\bf k}}}
\newcommand{\hp}{{\hat{\bf p}}}
\newcommand{\hx}{{\hat{\bf x}}}

\newcommand{\hq}{{\hat{\bf q}}}
\newcommand{\hy}{{\hat{\bf y}}}

\newcommand{\cL}{{\cal L}}

\newcommand{\lam}{{\lambda}}
\newcommand{\ga}{{\gamma}}

\newcommand{\be}{\begin{equation}}
\newcommand{\ee}{\end{equation}}
\newcommand{\bay}{\begin{eqnarray}}
\newcommand{\eay}{\end{eqnarray}}

\begin{document}

\title
{\bf Three charged particles in the continuum. Astrophysical examples.}
\author{Belyaev$^1$ V.B., Levin$^{2,3}$ S.B., Yakovlev$^3$ S.L.}
\date{}
\maketitle
{\small
{\it
\noindent
$^1$ N.N Bogolyubov Laboratory of Theoretical Physics, Joint Institute for Nuclear Research, Dubna, Russia\\
$^2$ Deparment of Physics, Stockholm University, Stockholm, Sweden\\
$^3$  Department of Mathematical and Computational Physics, V.A. Fock Institute of Physics,
St Petersburg State University, St Petersburg, Russia\\
}
}
\abstract{ \noindent We suggest a new adiabatic approach for
description of  three charged particles in the continuum.
This approach is based on the
Coulomb-Fourier transformation (CFT) of three body Hamiltonian,
which allows to develop a scheme, alternative to Born-Oppenheimer one.
The approach appears as an expansion
of the kernels of corresponding integral transformations in terms
of
small mass-ratio parameter. To be specific, the results are presented
for the system $ppe$ in the continuum. The wave function of a such
system is compared with that one which is used for estimation of
the rate for triple reaction
$
p+p+e\rightarrow d+\nu,
$
which take place as a step of $pp$-cycle in the center of the Sun. The
problem of microscopic screening for this particular reaction is
discussed.}

\section{Introduction.}
An accurate treatment of three charged particles in the continuum
at low relative energies represents till now a very
difficult problem which is actual in many areas of physics.
Indeed, only some special cases with specific properties of the
system has been considered in the literature. In this respect one
should mention papers \cite{p1,p2} where the asymptotic solution
for the three-body wave function has been obtained for configurations
when all interparticle distances are much larger of the
characteristic size of the system. An alternative limiting case
considered in \cite{p3} corresponds to configurations where one Jacobi
coordinate is much larger than the other one. The
near threshold breakup of hydrogen  by proton (or electron) studied in \cite{p4}
is another example of approximative
solution of three body Schr\"odinger equation obtained so far for three charged particles in
the continuum.

The purpose of this paper is to develop a new adiabatic expansion
for three-body Hamiltonian  for the
system consisting of one light and two heavy particles. The
asymptotic behavior of the wave-function with respect to the
coordinate of light particle  will be presented in the
framework of this new adiabatic expansion.
We use the Coulomb-Fourier transform formalism proposed in
\cite{ALY} to make 
a unitary transformation of the Hamiltonian, which leads to a
convenient representation, where, for example, one
long-range interactions is eliminated. 
The known explicit form of
eigenfunctions allows us to construct the useful integral
representation of interaction potentials for transformed
Hamiltonian. The important feature of this representation for potentials is the
appearance of universal integral with the integrand containing the exponential factor,
proportional to square root of mass ratio $\tau \sim \sqrt{m_e/m_p}$
of light (electron) and heavy (proton) particles. The natural power series in $\tau$
of this integral then
generate power series representation of the transformed Hamiltonian what is the basis
for our new variant of adiabatic representation of the problem.
As the first stage, in this paper
we study the solution of the problem taking into account  terms up to $O(\tau^2)$ order.

The paper is organized as following. Section 2 plays the central role and includes
formulation of the problem, description of necessary portions of Coulomb-Fourier
transformation. Here we give as exact formulas for CF transformed Hamiltonian
for $ppe$ system as well as derive the expansion of matrix elements of the Hamiltonian and
wave function as power series in $\tau^2$. Some technical details concerning
evaluation of singular integrals from the main tex is placed in
Appendix.
Third section contains application of approximation for three-body wave function obtained
in Section 2 to description of some reactions of $pp$ cycle. Short conclusion summarizes
the paper.

In the paper we use units such that $\hbar=c=1$, for the unit electric charge
the symbol $e$ is used. Three-dimensional vectors are denoted by
$\bx,\by,\bk,\bp...$ and
its modules and respective unite vectors by $x,y,k,p...$ and by $\hx,\hy,\hk,\hp ...$ .
Sometimes we  combine pairs of three-dimensional vectors in six-dimensional ones as
$\bX= \{ \bx,\by\}$, $\bP=\{\bk,\bp\}$. The Hilbert space of functions depending on vectors
$\bX$ which in our paper play the role of configuration space vectors will be denoted as
${\cal H}$ and the Hilbert space ${\hat {\cal H}}$ will be associated with function depending on
momentum variables $\bP$.

\section{Adiabatic expansion for three-body Hamiltonian and solution}

The $ppe$ system where $p$ is proton and $e$ is electron with masses $m_p$ and $m_e$
$(m_p  \gg m_e)$ respectively is considered. Number 1 we assign to electron whereas 2 and 3
to protons.
The Hamiltonian of the system in the center of mass frame using
mass-renormalized Jacobi coordinates
can be written in the form
$$
H= -\Delta_{\bx_1}-\Delta_{\by_1}+ v_s(x_1)+
n_1/x_1 + n_2/x_2 + n_3/x_3.
$$
Here $V_s(x_1)$ is a short-range potential describing strong $pp$ interaction. Mass-renormalized
charge factors $n_i$ are defined by the formulas
$n_i=e_j e_k \sqrt{2\mu_{jk}}$, where $e_1=-e$, $e_2=e_3=e$ are electron and protons charges and
$\mu_{ij}$
stands for reduced mass of a pair of particles $ij$, i.e.
$\mu_{ij}={m_i m_j/(m_i + m_j)}$. Introducing proton and electron masses into this
formula we get $\mu_{23}=m_p/2$, $\mu_{31}=m_e m_p/(m_e+m_p)$,
$\mu_{12}=\mu_{31}$.

Before proceed further, let us make three clarifying comments.
First,
throughout the paper we systematically use a convention that indices of any pair of
particles $ij$ are considered as a part of triad $ijk$ which itself is a cyclic
permutation of $1,2,3$.
Second, let us recall that mass-renormalized Jacobi set $\bx_i$, $\by_i$ is defined in such
a way that the vector $\bx_i$ up to the factor $\sqrt{2\mu_{jk}}$ is proportional to
the relative position vector of particles $j$ and $k$ and the vector $\by_i$ is the position vector
of particle $i$ with respect to the center of mass of corresponding two-body sybsystem.
There are three
possible sets $\bx_i$, $\by_i$, $i=1,2,3$ and different sets are related to each other by
kinematic rotation relations
\begin{eqnarray}
\bx_{i} &=& c_{ij} \bx_{j} + s_{ij} \by_{j}\nonumber
 \\
\by_{i}&=& -s_{ij} \bx_{i} + c_{ij} \by_{j}
\label{rot}
\end{eqnarray}
with coefficients being defined in terms of particle masses by the formulas
\be
c_{ij}=-\sqrt{\frac{m_i m_j}{(m_i+m_k)(m_j+m_k)}},
\label{Cij}
\ee
$s_{ij}=\epsilon_{ijk}\sqrt{1-c^2_{ij}}$, where
$\epsilon_{ijk}$ is fully antisymmetric tensor normalized as $\epsilon_{123}=1$. Third, in all mass
factors we keep general situation, i.e. not neglecting $m_e$ with respect to $m_p$,
making sometimes simplifications for illustrative purposes, as for example
$
\mu_{31}=m_e m_p/(m_e+m_p)=m_e(1+O(m_e/m_p)).
$

Now let us turn to the solution of the  Schr\"odinger equation
\be
H\Psi = E\Psi
\label{Schroe}
\ee
for three particles in the continuum $(E>0)$. To this end  we will construct
a special representation.
As the basis of this representation we take the eigenfunctions $\Psi_{c0}(\bX,\bP)$,
i.e. $H_{c0}\Psi_{c0}=\bP^2\Psi_{c0}$, of the operator
$$
H_{c0}=-\Delta_{\bx_1}-\Delta_{\by_1}+ n_1/ x_1
$$
with repulsive $(n_1>0)$ Coulomb potential. It is obvious that this eigenfunctions
have the form
$$
\Psi_{c0}(\bX,\bP)= \psi^c_{\bk_1}(\bx_1)\psi^0_{\bp_1}(\by_1).
$$
Here  $\psi^0_{\bp_1}(\by_1)=\frac{1}{(2\pi)^{3/2}}
e^{i\langle \bp_1,\by_1\rangle}$
is normalized plane wave and
\be
\psi^c_{\bk_1}(\bx_1)=\frac{1}{(2\pi)^{3/2}} e^{i\langle \bk_1,\bx_1\rangle}
e^{-\pi\gamma_{1}/2}\Gamma(1+i\gamma_{1})
\Phi(-i\gamma_{1}, 1, ik_{1}\xi_{1})
\label{psi Coul x}
\ee
is the normalized Coulomb wave function. The standard notations
for Sommerfeled parameter $\gamma_1=n_1/2k_1$,  parabolic
coordinate $\xi_1= x_1-\langle \bx_1,\hk_1 \rangle$, Gamma
function $\Gamma(z)$ and Confluent Hypergeometric function
$\Phi(a,c,z)$ have been used.

The representation which we call Coulomb-Fourier (CF) one is generated by the transform
\cite{ALY}
\be
G(\bX)= {\cal F}_c{\hat G}=\int d\bP \, \Psi_{c0}(\bX,\bP){\hat G}(\bP).
\label{CFtrans}
\ee
The integral operator ${\cal F}_c$ transforms the Hilbert space $\hat {\cal H}$ of functions
depending on momentum variables
$\bP$  into Hilbert space ${\cal H}$
of functions depending on coordinates.
Note that in the limiting case $n_1=0$ the ${\cal F}_c$ operator is reduced
to the standard inverse Fourier transform which connects conventional momentum space and
configuration space representations. The Hilbert space adjoint ${\cal F}^{\dag}_c$ acts from
${\cal H}$ to $\hat {\cal H}$ and for the pair ${\cal F}_c$ and ${\cal F}^{\dag}_c$
the unitarity properties hold
$$
{\cal F}^{\dag}_c {\cal F}_c = I_{\hat {\cal H}} , \ \ {\cal F}_c {\cal F}^{\dag}_c = I_{{\cal H}},
$$
which are just the operator form of the orthogonality and completeness of
eigenfunctions of the Hamiltonian $H_{c0}$.

In the Schr\"odinger equation in CF representation described above
\be
{\hat H}{\hat \Psi} =   E {\hat \Psi},
\label{CFSchroedinger}
\ee
the CF transformed Hamiltonian ${\hat H}= {\cal F}^{\dag}_c H {\cal F}_c$
appears now as integral operator with the kernel
(matrix elements)
\be
{\hat H}(\bP,\bP')= \langle \psi^0_{\bp_1}\psi^c_{\bk_1}
|H|
\psi^0_{\bp'_1}\psi^c_{\bk'_1}\rangle=
\label{hatH}
\ee
$$
(\bk_1^2+\bp_1^2)\delta(\bk_1-\bk'_1)\delta(\bp_1-\bp'_1) +
{\hat v}_s(\bk_1,\bk'_1)\delta(\bp_1-\bp'_1) +
W_2(\bP,\bP') + W_3(\bP,\bP')
$$
operating on CF-transformed wave function ${\hat \Psi}(\bP)$. Here the first
term corresponds
to kinetic energy operator ${\hat H}_0=\bk_1^2+\bp_1^2$, in the second term
${\hat v}_s$ stands for the CF-transformed short-range $pp$ interaction potential
$$
{\hat v}_s(\bk_1,\bk'_1) =\langle \psi^c_{\bk_1}|v_s|\psi^c_{\bk'_1}\rangle =
\int d\bx_1 \, \psi^{{c}*}_{\bk_1}(\bx_1)v_s(\bx_1)\psi^{c}_{\bk'_1}(\bx_1)
$$
where $*$ means the complex conjugation, and $W_{j}$ are Coulomb potentials
$n_{j}/x_{j}$ in CF representation. Let us notice,
that the contribution from $n_1/x_1$ potential  has been eliminated by CF transform.
The functions $W_{j}(\bP,\bP')$, $j=2,3$ have the following structure
$$
W_{j}(\bP,\bP')= |s_{j1}|^{-3}{\hat v}^{c}_{j}(s_{j1}^{-1}(\bp-\bp')){\cal L}_{j}(\bP,\bP'),
$$
where
$$
{\hat v}^{c}_{j}(\bq)=\frac{1}{2\pi^2}\frac{n_j}{|\bq|^2}
$$
is  the familiar Fourier transform of Coulomb potential $n_j/x_j$ and the functions
${\cal L}_{j}(\bP,\bP')$, $j=2,3$ are given by the integrals
\be
\cL_j(\bP,\bP')=\lim_{\lambda \to +0}\int d\bx_1\, e^{i\tau_j
\langle \bx_1,\bp-\bp'\rangle
-\lambda |\bx_1|}
\psi^{c*}_{\bk_1}(\bx_1)\psi^{c}_{\bk'_1}(\bx_1).
\label{Lj}
\ee
The parameters  $\tau_j$, $j=2,3$, have the kinematical origin
and are represented in terms of kinematic rotation matrix elements (\ref{rot}) as
$$
\tau_{j}= c_{j1}/s_{j1}.
$$

Noting that $\tau_3=-\tau_2$ which is the consequence of equality of heavy
particles (protons) masses and using definitions
(\ref{Cij}) and the fact that $m_e \ll m_p$ we get
$$
\tau_2 = \sqrt{m_e/2m_p}(1+O(m_e/m_p)),
$$
what shows that $\tau_j$ are small.
This allows us to expand the exponential factor in the integrand of
(\ref{Lj})
and obtain  in general case the expression
\be
\cL_j(\bP,\bP')=\delta(\bk_1-\bk'_1) +
\label{Lseries}
\ee
$$
\frac{i\tau_j}{1!}L^{(1)}(\bP,\bP')+
\frac{(i\tau_j)^2}{2!}L^{(2)}(\bP,\bP') +
\frac{(i\tau_j)^3}{3!}L^{(3)}(\bP,\bP') + ...
$$
Here $L^{(l)}(\bP,\bP')$  are integrals
\be
L^{(l)}(\bP,\bP')=\lim_{\lambda \to +0} \int d\bx_1 \, e^{-\lambda |\bx_1|}
{\psi^{c*}_{\bk_1}}(\bx_1)
\langle \bx_1,\bp_1-\bp'_1\rangle^{l}
\psi^{c}_{\bk'_1}(\bx_1)
\label{Lk}
\ee
which are independent on $j$. This last fact and the property $\tau_3=-\tau_2$
leads us to the following expansion for the sum of the CF transformed Coulomb potentials
$W_2+W_3$, which contains only even power terms
\be
W_2(\bP,\bP')+W_3(\bP,\bP')={\hat v}^{c}_{eff}(\bp_1,\bp'_1)\times
\label{W2+W3}
\ee
$$
\left\{ \delta(\bk_1-\bk'_1) -\frac{\tau^2}{2!}L^{(2)}(\bP,\bP')+
\frac{\tau^4}{4!}L^{(4)}(\bP,\bP')-...\right\},
$$
where we have introduced parameter $\tau=|\tau_2|$.
The quantity  ${\hat v}^{c}_{eff}(\bp_1,\bp'_1)$ stands for the Coulomb potential
corresponding to interaction between electron
and effective particle with the charge $2e$ and the mass $2m_p$ and has the form
\be
{\hat v}^{c}_{eff}(\bp_1,\bp'_1) = \frac{1}{2\pi^2}\frac{n_{eff}}{|\bp-\bp'|^2}
\label{effCoul}
\ee
with $n_{eff} = -2e^2\sqrt{2m_e} \sim
-2e^2\sqrt{2\frac{m_e 2 m_p}{m_e+2m_p}}$. 

Inserting formula (\ref{W2+W3}) into (\ref{hatH}) we arrive
at the representation of the CF-transformed Hamiltonian
$\hat H$ which plays a central role in the solution of the problem
\be
{\hat H} = {\hat H}_0 + {\hat V}_{s} + {\hat V}^{c}_{eff} + \tau^2{\hat W}.
\label{Ht2W}
\ee
The kernels of operators involved in (\ref{Ht2W}) read
$$
{\hat H}_0(\bP,\bP')=\left( \bk_1^2 + \bp_1^2 \right) \delta(\bP-\bP'),
$$
$$
{\hat V}_{s}(\bP,\bP')= {\hat v}_s(\bk_1,\bk'_1)\delta(\bp_1-\bp'_1),
$$

$$
{\hat V}^{c}_{eff}(\bP,\bP')={\hat v}^{c}_{eff}(\bp_1,\bp'_1)\delta(\bk_1-\bk'_1)
$$
and
\be
{\hat W}(\bP,\bP')= {\hat W}^{(0)}(\bP,\bP')-\tau^{2}{\hat W}^{(2)}(\bP,\bP')
+\tau^{4}{\hat W}^{(4)}(\bP,\bP')- ...,
\label{hat W}
\ee
\be
{\hat W}^{(l)}(\bP,\bP')=
{\hat v}^{c}_{eff}(\bp_1,\bp'_1)
\frac{1}{l!}L^{(l)}(\bP,\bP')
\label{hat W l}
\ee
where in the last case we have factored out the small parameter $\tau^2$ to show explicitly that the
last term in (\ref{Ht2W}) is as small as $\tau^2$.

The structure of the Hamiltonian (\ref{Ht2W}) suggests now the natural perturbative scheme
for solution of
Schr\"odinger equation (\ref{CFSchroedinger}). Let us represent the  wave-function ${\hat \Psi}$
as power series in $\tau^2$
\be
{\hat \Psi} = {\hat \Psi}_0 + \tau^2 {\hat \Psi}_2 + \tau^4 {\hat \Psi}_4 +...
\label{Psi_power}
\ee
then inserting (\ref{Psi_power}) into (\ref{CFSchroedinger}) one immediately gets the recursive
set of equations for ${\hat \Psi}_k$, i.e.
\be
\left( {\hat H}_0 + {\hat V}_s + {\hat V}^{c}_{eff}\right) {\hat \Psi}_0 = E{\hat \Psi}_0,
\label{t1}
\ee
\be
\left( {\hat H}_0 + {\hat V}_s + {\hat V}^{c}_{eff}\right) {\hat \Psi}_{2l}
= E{\hat \Psi}_{2l} - \sum_{s=0}^{l-1}(-1)^{l-s}{\hat W}^{(2l-2s)}{\hat \Psi}_{2s},\ \ \  l=1,2,3, ...
\label{tk}
\ee
The scheme (\ref{t1}, \ref{tk}) has a remarkable property, namely, the solution of the three-body
problem
in framework of this scheme can be obtained in terms of solutions of two-body problems. Indeed,
equation
(\ref{t1}) allows the separation of variables, so that its solution is reduced to the solution of
respective
two-body equations, at the same time the solution of inhomogeneous equations (\ref{tk}) can be
obtained
in terms of Green's function of the operator ${\hat H}_0 + {\hat V}_s + {\hat V}^{c}_{eff}$ which
can be constructed from two-body Green's functions due
to separability of variables.

For the specific case
of three particles in the continuum we are considering the solution scheme outlined above
yields the following results. The solution of the first equation (\ref{t1}) reads
\be
{\hat \Psi}_0(\bP,\bP^{in})={\hat \psi}_{\bk^{in}_1}^{+}(\bk_1)
{\hat \psi}^{ce}_{\bp^{in}_1}(\bp_1),
\label{psi1}
\ee
where initial state momentums $\bk^{in}_1$ and $\bp^{in}_1$ are related to the energy $E$
by the formula
${\bP^{in}}^{2}={{\bk^{in}_1}^2} +{{\bp^{in}_1}^2} = E$.
The function ${\hat \psi}^{ce}_{\bp^{in}_1}(\bp_1)$ is the
momentum space Coulomb
wave function corresponding to the potential ${\hat v}^{c}_{eff}(\bp_1,\bp'_1)$. The inverse
Fourier transform of $\psi^{ce}_{\bp^{in}_1}(\bp_1)$ we will denote by
$\psi^{ce}_{\bp^{in}_1}(\by_1)$ which explicit form can be obtained from (\ref{psi Coul x})
when $n=n_{eff}$ and replacing $\bk^{in}_1,\bx_1$ by $\bp^{in}_1,\by_{1}$, respectively. The term
 ${\hat \psi}_{\bk^{in}_1}^{+}(\bk_1)$ is the scattering solution to the two-body Schr\"odinger
equation with the potential ${\hat v}_s(\bk_1,\bk'_1)$ and is conventionally represented as the
solution of the Lippmann-Schwinger integral equation
\be
{\hat \psi}_{\bk^{in}}^{+}(\bk) = \delta(\bk-\bk^{in}) - \frac{1}{\bk^2-{\bk^{in}}^2-i0}
\int d\bq \, {\hat v}_s(\bk,\bq){\hat \psi}_{\bk^{in}}^{+}(\bq).
\label{LS}
\ee
The solutions of inhomogeneous equations (\ref{tk}) are given by recursive formulas starting
from ${\hat \Psi}_0(\bP,\bP^{in})$
\be
{\hat \Psi}_{2l} = -{\hat G}_{s, eff}(E+i0)\sum_{s=0}^{l-1}(-1)^{l-s}{\hat W}^{(2l-2s)} {\hat \Psi}_{2s}.
\label{psik}
\ee
Here the kernel of the operator
${\hat G}_{s, eff}(z)= \left( {\hat H}_0 + {\hat V}_s + {\hat V}^{c}_{eff}-z\right)^{-1}$
is represented through
two-body Green's functions ${\hat g}_{s}$ for potential ${\hat v}_s$ and ${\hat g}^{c}_{eff}$
for potential ${\hat v}^{c}_{eff}$
by the counter integral
$$
{\hat G}_{s, eff}(\bP,\bP', z) = \frac{1}{2i\pi } \oint_{C}d \zeta\,
 {\hat g}_{s}(\bk_1,\bk'_1,\zeta ){\hat g}^{c}_{eff}(\bp_1,\bp'_1,z-\zeta)
$$
with counter $C$ encircling the cut of ${\hat g}_{s}$
in
anticlockwise direction.

So that, we have constructed the formal solution to the CF transformed
Schr\"odinger equation (\ref{CFSchroedinger})
for the system $ppe$ in the continuum. The configuration space wave function which obeys (\ref{Schroe})
can be obtained from ${\hat \Psi}$ by CF transform (\ref{CFtrans})
\be
\Psi(\bX,\bP^{in}) = \int d\bP \Psi_{c0}(\bX,\bP){\hat \Psi}(\bP,\bP^{in}).
\label{XSolution}
\ee
Now one can see, that 
the structure of our solution (\ref{Psi_power}) and respective series in $\tau^2$
for $\Psi(\bX,\bP^{in})$ generated from (\ref{XSolution}) by (\ref{Psi_power}) and the structure
of the representation for the Hamiltonian (\ref{Ht2W}) outline the framework of our approach as
an alternative to Born-Oppenheimer one.
It is worth mentioning here, that the formalism given above is rather general and with minor
evident
modifications is applicable for the three charged particle systems for the case of different
masses when
the mass of one particle is significantly smaller than the masses of others.

Before proceed further, let us give some explicit formulas
for  approximation to the wave function $\Psi(\bX,\bP^{in})$ generated by  our complete
formal solution  which we will use in the next Section discussing some astrophysical reactions.
Introducing (\ref{psi1}) and (\ref{psik}) into (\ref{XSolution}) we get
\be
\Psi(\bX,\bP^{in}) = \psi^{+}_{\bk^{in}_1}(\bx_1)\psi^{ce}_{\bp^{in}_{1}}(\by_1) +
\tau^2 \Psi_2(\bX,\bP^{in}) + O(\tau^4)
\label{pprox}
\ee
where
$$
\psi^{+}_{\bk^{in}_1}(\bx_1) = \int d\bk_1 \psi^{c}_{\bk_1}(\bx_1)
{\hat \psi}^{+}_{\bk^{in}_1}(\bk_1)
$$
and
$\Psi_2(\bX,\bP^{in})$ is given by transform (\ref{XSolution}) of ${\hat \Psi}_2$ calculated through
${\hat \Psi}_{0}$  from (\ref{psi1}) by formula
\be
{\hat \Psi}_{2}=-{\hat G}_{e,eff}(E+i0){\hat W}^{(2)}{\hat \Psi}_{0}.
\label{Psi 2}
\ee
The complete investigation of properties of the solutions to
(\ref{t1},\ref{tk}) is out of the scope of this paper and will be made elsewhere. Below in this
section,  we consider two points which plays the key role for the formalism, namely
the singular structure of operator $\hat W$ and  the structure of
correction term ${\hat \Psi}_{2}$ (and consequently $\Psi_{2}$)
 which possesses  the most important properties specific for all correction terms ${\hat \Psi}_{2l}$.

The kernels of the operators ${\hat W}^{(l)}$ in (\ref{hat W l}) are given in terms of singular integrals
(\ref{Lk}).
These integrals obviously can be computed in terms of distributions (generalized functions)
as it could be  seen from (\ref{Lk}) in the trivial case $l=0$ which yields  $L^{(0)}(\bP,\bP')=
\delta(\bk_1-\bk'_1)$. The general case of arbitrary $l$ in (\ref{Lk}) is considered in the
Appendix where it is shown that the most singular part of the integral
$L^{l}(\bP,\bP')$
has the form
\be
L^{(l)}(\bP,\bP') = \delta(\hk,\hk')\delta^{(l)}(k-k')\langle \hk,\bp-\bp'\rangle^{l}
{\cal L}^{(l)}(k,k').
\label{Lm}
\ee
Here and in what follows we omit subscript 1 from definition of momentums and coordinates
using, for instance, $\bk$ instead of $\bk_1$ and so ones.
Delta-function $\delta(\hk,\hk')$ on unit sphere $S^2=\{\bk: k=1\}$
 and $l$-th derivative
of delta-function $\delta^{(l)}(k-k')$ are defined by
$$
\int_{S^2} d\hk\, \delta(\hk,\hk')g(\hk') = f(\hk),
$$
$$
\int_{-\infty}^{\infty} dk'\, \delta^{(l)}(k-k')g(k) = (-1)^{l}g^{(l)}(k).
$$
The function ${\cal L}^{(l)}(k,k')$ being a smooth function of $k$ and $k'$
for $l$-even has the form
$$
{\cal L}^{(l)}(k,k') = \frac{1}{kk'}e^{-i(\sigma(k)-\sigma(k'))}
\frac{\sinh{\frac{\pi(\gamma-\gamma')}{2}}}
{\frac{\pi(\gamma-\gamma')}{2}} \times
$$
$$
\Re
\left[
e^{i(\sigma_{0}(k)-\sigma_{0}(k'))+i\frac{\pi l}{2}} (2k)^{-i\gamma}(2k')^{i\gamma'}
\Gamma(1-i(\gamma - \gamma'))
\right] .
$$

Above formulas for $L^{(l)}(\bP,\bP')$ can be used to compute the action of ${\hat W}^{(2)}$ operator on
${\hat \Psi}_0$
\be
{\hat W}^{(2)}{\hat \Psi}_{0}(\bP, \bP^{in})=
\frac{1}{2}I_{1}(\bk,\bk^{in})I_{2}(\bp,\bp^{in})
\label{W2Psi0}
\ee
where
$$
I_{1}(\bk,\bk^{in})= \int d\bk'\, \delta(\hk,\hk')\delta^{(2)}(k-k'){\cal L}^{(2)}(k,k')
{\hat \psi}^{+}_{\bk^{in}}(\bk')
$$
and
$$
I_{2}(\bp,\bp^{in})= \int d\bp'\, {\hat v}^{c}_{eff}(\bp,\bp')\langle \hk,\bp-\bp'
\rangle^{2}{\hat \psi}^{ce}_{\bp^{in}}(\bp').
$$
Both integrals $I_{j}$ are singular distributions. To make its structure more transparent let us find
its  most singular parts. For the first integral it means leaving in the integrand the most singular part of
${\hat \psi}^{+}_{\bk^{in}}(\bk')$ from (\ref{LS}), i.e. $\delta(\bk-\bk^{in})$ which yields
\be
I_1(\bk,\bk^{in})= \delta(\hk,\hk^{in})\delta^{(2)}(k-k^{in}){\cal L}^{(2)}(k,k^{in}).
\label{I 1}
\ee
For the second integral, it is useful to make a linear change of variables to get
$$
I_2(\bp,\bp^{in}) = \frac{n_{eff}}{2\pi^2}\int d\bq\, \langle \hk,\hq \rangle^{2}
{\hat \psi}^{ce}_{\bp^{in}}(\bq+\bp)
$$
and then using Fourier transform for ${\hat \psi}^{ce}_{\bp^{in}}(\bp)$ rewrite this integral as
\be
I_2(\bp,\bp^{in}) =\frac{n_{eff}}{2\pi^2}\int d\by\,  D(\by,\hk) e^{-i\langle \bp,\by \rangle}
\psi^{ce}_{\bp^{in}}(\by).
\label{I2 D}
\ee
Here the function $D(\by,\hk)$ is given by
$$
D(\by, \hk)= \lim_{\lambda \to +0} \frac{1}{(2\pi)^{3/2}} \int d\bq\,
e^{-i\langle \bq,\by \rangle-\lambda q}\langle \hk,\hq\rangle^{2}.
$$
In the Appendix it is shown that the main singular part of $D(\by,\hk)$ is proportional to delta-function,
i.e.
$$
D(\by,\hk)= \frac{1}{3(2\pi)^{3/2}}\delta(\by).
$$
The latter gives for the integral $I_{2}(\bp,\bp^{in})$
\be
I_{2}(\bp,\bp^{in})=N_{eff}
\psi^{c}_{\bp^{in}}(0),
\label{I 2}
\ee
$$
N_{eff}=\frac{n_{eff}}{2\pi^2}\frac{1}{3(2\pi)^{3/2}}.
$$

Now, inserting formulas given above in (\ref{Psi 2}) we can represent the correction term
${\hat \Psi}_{2}$
as the integral
\be
{\hat \Psi}_{2}(\bP,\bP^{in})= -\frac{1}{4\pi i}
\int d\bP' \oint_{C} d\zeta \,
{\hat g}_{s}(\bk,\bk',\zeta){\hat g}^{c}_{eff}(\bp,\bp',E-\zeta +i0)\times
\label{Psi 2 I1 I2}
\ee
$$
I_{1}(\bk',\bk^{in}) I_{2}(\bp',\bp^{in}).
$$
This general formula can be simplified if we take instead of full Green's function
${\hat g}_{s}(\bk,\bk',\zeta)$ its main singular part which is Green's function of
the two-body kinetic energy operator
 $\delta(\bk-\bk')(k^2 -\zeta)^{-1}$. This case in fact has the particular
physical sense, since taking into account that we left only delta-function for ${\hat \psi}^{+}_{s}$
the resulting approximation is exactly equivalent to the neglect of the short-range potential $V_{s}$
from the very beginning.
Formula (\ref{Psi 2 I1 I2}) is transformed in this case to
\be
{\hat \Psi}_{2}(\bP,\bP^{in}) =
\label{Psi 2 m}
\ee
$$
-\frac{1}{2}N_{eff}
I_{1}(\bk,\bk^{in})\int d\bp' {\hat g}^{c}_{eff}(\bp,\bp',E-k^{2}+i0)
\psi^{ce}_{\bp^{in}}(0).
$$

The configuration space representation for $\Psi_{2}$ which can be obtained from formula
(\ref{XSolution}) is reduced now to the integral
$$
\Psi_{2}(\bX,\bP^{in})=
$$
$$
 -\frac{1}{2}N_{eff}\psi^{c}_{\bp^{in}}(0)
\int d\bk \, \psi^{ce}_{\bk}(\bx) I_{1}(\bk,\bk^{in}) g^{c}_{eff}(\by, 0,E-k^2+i0).
$$
Final form for this integral follows immediately from delta-functional factors of $I_1$ and reads
\be
\Psi_{2}(\bX,\bP^{in})=
-\frac{1}{2}N_{eff}\psi^{ce}_{\bp^{in}}(0) \times
\label{Psi 2 final}
\ee
$$
\left.
\frac{\partial^{2}}{\partial t^2}
\left[ t^{2}{\cal L}^{(2)}(t,k^{in})\psi^{c}_{t\hk^{in}}(\bx)
g^{c}_{eff}(\by, 0,E-t^2+i0)\right] \right|_{t=k^{in}}.
$$
This formula describes the correction term $\Psi_{2}$ for the $ppe$ system when strong $pp$
interaction is neglected and at the same time is approximation to the term $\Psi_2$ in the general case.

The formula (\ref{Psi 2 final}) is useful for constructing the coordinate asymptotics of
$\Psi_{2}(\bX,\bP^{in})$ as $y \to \infty$. One needs to use well known coordinate asymptotics
of Coulomb Green's function as $y \to \infty$ and $y'$ is bound
$$
g^{c}_{eff}(\by,\by',s^2+i0) \sim \frac{\exp{\{ i sy - i \frac{n_{eff}}{2s}\log 2sy\}}}
{4\pi y }
\psi^{c*}_{-s\hy}(\by')
$$
which can be found for instance in \cite{MF}. This asymptotic formula gives the following asymptotics
of $\Psi_{2}(\bX,\bP^{in})$
\newpage
\be
\Psi_{2}(\bX,\bP^{in}) \sim
\label{Psi 2 asymp}
\ee
$$
{\cal A}(\bx,\bk^{in},\bp^{in},\hy)
\frac{\exp{\{ i p^{in}y - i \frac{n_{eff}}{2p^{in}}\log 2p^{in}y \}}}
{4\pi y }
\left( 1+ O\left(y \frac{k^{in}}{p^{in}}\right)\right)
$$
where the amplitude ${\cal A}$ has the  explicit form
$$
{\cal A}(\bx,\bk^{in},\bp^{in},\hy)=
-\frac{1}{2}N_{eff}\psi^{c}_{\bp^{in}}(0)\times
$$
$$
\left.
\frac{\partial^{2}}{\partial t^2}
\left[ t^{2}{\cal L}^{(2)}(t,k^{in})\psi^{c}_{t\hk^{in}}(\bx)
\psi^{ce*}_{-\sqrt{E-t^2}\hy}(0)
\right] \right|_{t=k^{in}}
$$
Here by $O\left(y \frac{k^{in}}{p^{in}}\right)$ we have denoted  terms corresponding to
derivatives of exponential factor in (\ref{Psi 2 final}). The order of this terms shows the
range of validity of the asymptotics (\ref{Psi 2 asymp}), i.e. $y \frac{k^{in}}{p^{in}}$ has to be
small, what in terms of masses must be equivalent to the fact that $y \tau^{2}$ has to be small.
Let us emphasize that all treatment above devoted to consideration of three body charged particles at 
kinetic energies, comparable with corresponding potential energy, which means that Coulomb interactions
are essential. In this situation for the systems, consisting from heavy and light particles, one can develop 
the adiabatic description, which actually means smallnes of momentums of heavy and light particles ratio
$\frac{k^{in}}{p^{in}}$. This smallnes obviously appear due to the small mass ratio parameter $\tau$, introduced above.

Therefore in each fixed order of expansion in small parameter $\tau$ one should neglect also by all terms, 
proportional to the ratio $\frac{k^{in}}{p^{in}}$.


\section{Astrophysical Examples}

Now let us discuss ways of describing some reactions of the $pp$-cycle on the
Sun, which can be done on the ground of 3-body wave function given by
(\ref{pprox}). In
other words, we will consider situations when in the initial state the system
consists of three charged particles in the continuum and the mass of one of them is
much smaller than other masses.

The first example gives the reaction
\be
                p+p+e \rightarrow d+\nu
\label{dnu}
\ee
considered in \cite{Bahcal}. As it follows from the form of the main term in the
right hand side of (\ref{pprox}), with very good accuracy we
have separation of
the Jacobi coordinates in the wave function of the initial state for the reaction (\ref{dnu}).
This means that the
rate of three-body process (\ref{dnu}) can be expressed in terms of a
binary process
\be
               p+p\rightarrow d+e^+ +\nu
\label{denu}
\ee
This is just the main result of paper \cite{Bahcal}. Now it becomes clear that
the physical background of the above result from the point of view
of the few-body theory consists in two points:

a) the system has two heavy and one light particle
such that the parameter $\frac{m}{M}\ll 1$
and therefore one can neglect the second term in the
right-hand side of (\ref{pprox}).

b) heavy particles are slow
enough to neglect
higher partial waves in their relative motion. One should emphasize that
free "effective charge of the initial nuclear system"  $Z$, introduced
in \cite{Bahcal}, can now be fixed to value $Z=2$ which is supported by the structure of
(\ref{pprox}).

Let us consider another example of 3-body initial state
\be
            p+^7Be+e,
\label{pbe}
\ee
which can produce $^8$Be or $^7$Li nuclei via the following reactions
$$
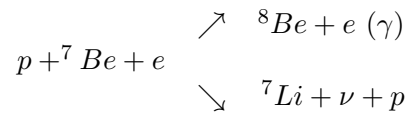
\begin{tabular}{lcr}
  & $\nearrow$ & $^8Be+e\ (\gamma)$ \\
$p+^7Be+e$ & & \\
 & $\searrow$ & $^7Li+\nu+p$ \\
\end{tabular}
$$

First, from the previous discussion one can see that due to different
masses of heavy particles in this case the contribution from the linear term
$\tau$ is nonzero in contrast to the $p+p+e$ system, and this contribution
should be estimated.

If the electron in state (\ref{pbe}) is in the continuum, then again
due to the
separation of Jacobi coordinate in the first term of (\ref{pprox})
the rate of the
proton capture from the initial three-body state (\ref{pbe}) can be expressed
via the rate
of the binary reaction $p+^7Be\rightarrow ^8B+\gamma$.

However, the rate of the electron capture from the initial
three-body state (\ref{pbe}),
as it follows from (\ref{pprox}), (modified for the state (\ref{pbe})),
will be defined by the
Coulomb wave function of the electron moving in the Coulomb
field with the charge
$Z=5$ instead of $Z=4$ for the capture from the two-body state $^7$Be+$e$.
This means that the production rate of $^7$Li from the three-body state (\ref{pbe})
cannot be expressed via the binary ($e$+$^7$Be$\rightarrow$$^7$Li+$\nu$)
reaction rate.
Roughly speaking, the ratio of these rates will be proportional
to the ratio of the corresponding electron Coulomb functions at energy
in the center of the Sun $E_s$. In other words,
$$
\frac{w_3}{w_2}\sim \left |\frac{\psi_c(0,E_s,Z=5)}{\psi_c(0,E_s,Z=4)}
\right|^2 \sim \frac54.
$$

Now let us discuss the problem of screening of the Coulomb interaction
between two protons for the system $p+p+e$. We restrict ourselves by lowest
order in the ratio $m_e/m_p$ for the three-body wave function, i.e. consider only
first term in the (\ref{pprox}). It is evident, that the screening effect
in this approximation appears due to the electronic wave function
$\psi_c(\bp,\by)$, where $\by=\sqrt{\frac{4m_em_p}{m_e+2m_p}}\left(\frac{\bR}{2}+\br\right)$,
$\bR$ being the distance between
protons and $\br$ being the distance between electron and one of the protons.
Taking the asymptotics of this function in the region where $R\gg r$, one can see
that the Coulomb phases of $pp$ wave function and electronic wave function can
cancel each other  for the specific configurations of initial momentums $\bk$ and
$\bp$ of three-body system under consideration. Hence the resulting motion
of two protons  in this configuration would be described by plane wave,
which means the total screening effect.

\section{Conclusion}

In conclusion we emphasize, that the CF-transformed three-body Hamiltonian
(\ref{Ht2W}) for the system of two heavy and one light particles can be used
for realization of adiabatic expansion which is  alternative to the
Born-Oppenheimer one.
This approach allows to treat screening effects on the microscopic level.
In astrophysical examples it was shown, that in the lowest order of small parameter $\tau$,
it is possible for some reactions only express rates of 3-body processes in terms of binary ones.

\newpage

\section{Appendix}
In this section we give a brief description of evaluation stages of integrals $L^{(l)}(\bP,\bP')$
and $D(\by,\hk)$ which are defined as
$$
L^{(l)}(\bP,\bP')=\lim_{\lambda \to +0}L^{(l)}(\bP,\bP',\lambda),
$$
\be
L^{(l)}(\bP,\bP',\lambda)=
\int d\bx_1 \, e^{-\lambda x_1}
{\psi^{c*}_{\bk_1}}(\bx_1)
\langle \bx_1,\bp_1-\bp'_1\rangle^{l}
\psi^{c}_{\bk'_1}(\bx_1)
\label{ALk}
\ee
and
$$
D(\by, \hk)= \lim_{\lambda \to +0}D(\by, \hk,\lambda),
$$
\be
D(\by, \hk,\lambda)=  \frac{1}{(2\pi)^{3/2}} \int d\bq\,
e^{-\lambda q-i\langle \bq,\by \rangle}\langle \hk,\hq\rangle^{2}.
\label{A D yk}
\ee
The technical tool we use for calculations of prelimiting integrals (\ref{ALk}, \ref{A D yk})
is so called weak asymptotics. For Coulomb wave function $\psi^{c}_{\bk}(\bx)$ as $y \to \infty$
\cite{MF} it reads
\be
\psi^{c}_{\bk}(\bx) \sim \frac{i}{(2\pi)^{1/2}k}\times
\label{A psi weak}
\ee
$$
\left\{
\delta(\hx,-\hk)\frac{\exp{(-ikx -iw_0(x,k))}}{x} - s_{c}(k,\hx,\hk)
\frac{\exp{(ikx+ iw_0(x,k))}}{x},
\right\}
$$
where $w_0(x,k)=-\ga \log 2kx$, $\ga=n/2k$ and $s_{c}(k,\hx,\hk)$ is the Coulomb s-matrix
$$
s_{c}(k,\hx,\hk)= \frac{2^{1+2i\ga}\; \ga
}{2i\pi}\frac{e^{ 2i\sigma_{0}}} {|\hx-\hk|^{2+i2\ga}},
$$
$\sigma_0(k)=\arg \Gamma(1+i\gamma)$.
If $n=0$ this formula is reduced to the weak asymptotics of plane wave \cite{Peterkop}
\be
\frac{1}{(2\pi)^{3/2}} e^{i\langle \bk,\bx\rangle}\sim
\frac{i}{(2\pi)^{1/2}k}
\left\{
\delta(\hx,-\hk)\frac{\exp{(-ikx)}}{x} - \delta(\hx,\hk)\frac{\exp{(ikx)}}{x}
\right\}.
\label{A pw wa}
\ee
The last key formula we need to compute above integrals is
$$
\int d \hk \, s_{{c}}(k,\hx,\hk) g(\hk)
=
e^{2i\sigma_{0}} g(\hx) + \frac{2^{i\ga}\;\ga e^{2i\sigma_{0}}
}{2i\pi} \int_0^2 dt \frac{G(t)-G(0)}{t^{1+i\ga}} ,
$$
with
$$
G(0)=2\pi g(\hx),\; G(t)=\int_0^{2\pi}d\phi \,
g(t,\phi).
$$
In these equations, $\theta$ and $ \phi$ are spherical angles and $t=1-\cos \theta$. In the sense of
distribution it means that
\be
s_{{c}}(k,\hx,\hk) = e^{2i\sigma_{0}}\delta(\bx,\bk) + ...
\label{A s_c main}
\ee

The integral from (\ref{ALk}) for $\lambda \ne 0$ can be calculated by following procedure, inserting
(\ref{A psi weak}) into the integral, leaving the main singular part of $s_{c}$ from (\ref{A s_c main}),
and using the formula \cite{RG}
$$
\int_{0}^{\infty} dx \, x^{ia} \,e^{\pm i t x -\lam x} =e^{\pm
i\pi/2 \mp \pi a/2}\frac{\Gamma(1+ia)} {(t \pm i\lam)^{1+ia}}.
$$
one arrives at
$$
L^{(l)}(\bP,\bP',\lambda)=\frac{1}{2\pi kk'}\delta(\hk,\hk')e^{-i\sigma_{0}(k)+\sigma_{0}(k')}\times
$$
$$\left[
\langle -\hk,\bp-\bp'\rangle^{l}B_l(k,k'\lambda) +
\langle \hk,\bp-\bp'\rangle^{l}B^{*}_l(k,k'\lambda)
\right],
$$
where
$$
B_l(k,k'\lambda)=
e^{i\sigma_{0}(k)-i\sigma_{0}(k')}
(2k)^{-i\ga}(2k')^{i\ga}e^{\frac{i\pi(l+1)}{2}}
\frac{\Gamma(1+l-i(\ga-\ga'))}{(k-k'+i\lambda)^{1+l-i(\ga-\ga')}}.
$$
It remains to evaluate the $\lambda \to 0$ limit, which can be done by means of the following
representation
which one can  verify by straightforward calculations
$$
\lim_{\lambda \to +0} \int^{\infty}_{-\infty} \frac{dt\, g(t)}{(t\pm i\lambda)^{n+i\mu}}=
\frac{\Gamma(1+i\mu)}{\Gamma(n+i\mu)}
\left\{
i\mu^{-1}g^{(n-1)}(0)\left[1-e^{\pm \pi \mu}\right] +
\right.
$$
$$
\left.
\int^{1}_{-1}\frac{dt\, [g^{(n-1)}(t)-g^{(n-1)}(0)]}{(t \pm i0)^{1+i\mu}} +
\left(\int_{-\infty}^{-1}+\int_{1}^{\infty}\right)
\frac{dt\, g^{(n-1)}(t)}{(t \pm i0)^{1+i\mu}}
\right\}.
$$
Using symbolic notations and leaving explicitly only main singular part
$$
\frac{1}{(t\pm i0)^{n+i\mu}} =
\frac{\Gamma(1+i\mu)}{\Gamma(n+i\mu)}i\mu^{-1}(-1)^{n-1}\delta^{(n-1)}(t)
\left[1-e^{\pm \pi \mu}\right] +...
$$
we arrive finally at the formula (\ref{Lm}) given in the main text.

By very similar way the integral (\ref{A D yk}) can be evaluated, i.e. usage of (\ref{A pw wa})which is
at the same time the asymptotics of plane wave as $q \to \infty$
helps to calculate the integral over $\hq$ and then the radial integral over $q$ gives
$$
D(\by,\hk,\lambda)= \frac{1}{(2\pi)^{3/2}}\frac{8\pi\lambda \langle \hy,\hk\rangle^{2}}
{(y^2+\lambda^2)^2}.
$$
The $\lambda \to +0$ limit is now straightforward
$$
\lim_{\lambda \to +0} D(\by,\hk,\lambda) = \frac{1}{3(2\pi)^{3/2}}\delta(\by).
$$


\begin{thebibliography}{99}

\bibitem{p1} R.K. Peterkop, Zh. Eksp. Teor. Fiz, {\bf 43}, 616 (1962)
(in russian) [Sov. Phys. JETP {\bf 14}, 1377 (1962)].

\bibitem{p2} S.P. Merkuriev, Theor. Math. Phys., {\bf 32}, 680 (1977),
M. Brauner, J.S. Briggs and H.J. Klar,  J. Phys. B, {\bf 22}, 2265 (1989).

\bibitem{p3} E.O. Alt, A.M. Mukhamedzhanov,  JETP Lett.,
{\bf 56}, 435 (1992),  Phys. Rev. A, {\bf 47}, 2004 (1993);
Y.E. Kim, A.L. Zubarev, Phys. Rev. A, {\bf 56}, 521 (1997).

\bibitem{p4} J.H. Macek, S.Yu. Ovchinnikov,
 Phys. Rev. A, {\bf 54}, 1 (1996);
M.Yu.Kuchiev and V.N.Ostrovsky, Phys. Rev. A, {\bf 58}, 321 (1998).

\bibitem{ALY} E.O. Alt, S.B. Levin and  S.L. Yakovlev, {\it Coulomb-Fourier
transformation: a novel approach to three-body scattering with
charged particles}. (submitted to  Phys. Rev. C); 
E.O. Alt, S.B. Levin and  S.L. Yakovlev, Few-Body Systems Suppl. {\bf 14}, 221 (2003); 
E.O.Alt, Few-Body Systems Suppl. {\bf 14}, 179 (2003); E.O. Alt, S.B. Levin and  S.L. Yakovlev,
Book of Abstracts, Few Body 17, Durham 2003, p. 283, 287;
Belyaev V.B., Levin S.L., Yakovlev S.L., ibid, p. 346.

\bibitem{Bahcal} J.N. Bahcall, R.M. May,
Astrophysical Journal, {\bf 155}, 501, (1969).


\bibitem{MF} L.D. Faddeev, S.P. Merkuriev,
{\it Quantum Scattering Theory for Several Particle Systems,}
(Kluwer, Dordrecht, 1993).

\bibitem{LAY} S.B. Levin, E.O. Alt and S.L. Yakovlev, {\it Integral Representation
for the Two-Body Coulomb Wave Function},
in: Selected topics in theoretical physics and astrophysics:
collection of papers dedicated to Vladimir B. Belyaev on the occesion of his
birthday. - Dubna: JINR, 2003,  167 p.; ISBN 5-9530-0022-7.

\bibitem{Peterkop} R.K. Peterkop, {\it Theory of Ionization of Atoms by Electron Impact}
(Colorado Associated University Press, Buolder, 1977).

\bibitem{RG} I. S. Gradshteyn and I. M. Ryzhik, {\it Table
of Integrals, Series, and Products} (Academic Press, San Diego,
1980).
\end{thebibliography}
\end{document}